# Conformable Fractional Isothermal Gas Spheres

Eltayeb A. Yousif[1,2], Ahmed M. A. Adam[1,3], Abaker A. Hassaballa[1,4], and Mohamed I. Nouh[5]

[1] Department of Mathematics, Faculty of Science, Northern Border University, Arar 91431, Saudi Arabia
[2] Department of Applied Mathematics, Faculty of Mathematical Sciences, University of Khartoum, Khartoum 11111, Sudan
[3] Faculty of Engineering, Alzaiem Alazhari University, Khartoum North 13311, Sudan
[4] Department of Mathematics, College of Applied & Industrial Sciences, Bahri University, Khartoum, Sudan
[5] Astronomy Department, National Research Institute of Astronomy and Geophysics (NRIAG), 11421 Helwan, Cairo, Egypt

e-mail: abdo_nouh@hotmail.com

**Abstract:**

The isothermal gas sphere is well known as a powerful tool to model many problems in astrophysics, physics, chemistry, and engineering. This singular differential equation has not an exact solution and solved only by numerical and approximate methods. In the present paper and within the framework of the Newtonian hydrostatic equilibrium, we have developed general analytical formulations for the fractional isothermal gas sphere. To obtain analytical expressions for mass, radius, and density, besides the fractional isothermal gas sphere, we used the conformable fractional calculus. Using the series expansion method, we obtained a general recurrences relation, which allows us to determine the series coefficients. The comparison of the series solution with the numerical ones for the fractional parameter $\alpha = 1$ is good for dimensional parameters up to $x \simeq 3.2$, beyond this value, the series diverges. We applied a combination of Euler-Abel and Pade techniques to accelerate the series, therefore accelerated series converge to the numerical desired value. We analyzed some physical parameters of a typical model of the neutron stars such as the mass-radius relation, density, and pressure ratio for different models. We found that the current models of the conformable neutron stars had smaller volumes and masses than both stars in the context of modified Rienmann-Liouville derivatives as well as the integer one.

1. Introduction

In mathematical physics and astrophysics, unusual anomalies and studies of singular problems with intrinsic significance within second order ordinary differential equations have been of great importance and many mathematicians and physicists have attracted interest. One of these equations is the second type of the singular Lane-Emden equation describing the isothermal gas sphere which is useful in many astrophysical fields, such as planetary interfaces, star clusters, galaxies as well as galactic clusters, Binny and Tremaine (1987), Horedt (2004).

Many methods have been introduced to solve the isothermal gas sphere (it may also be called the second type Lane-Emden differential equation) in its integer form, Natarajan and Lynden-Bell (1997), Roxburgh and Stockman (1999), Hunter (2001), Nouh (2004), Ibrahim and Dares (2008), Momani and Ibrahim (2008), Chowdhury (2009).

Over the past 20 years, interest in the study of fractional differential equations has grown in many fields of science and engineering: mathematics, chemistry, optics, plasma, fluid dynamics. Applications of the fractional calculus in astrophysics attract more attention in the recent past, Stanislavsky (2010). By selecting a power-law weight function in the cosmology of fractional action, Momeni and Rashid (2012) introduced the dark energy model with relevant cosmological parameters. El-Nabulsi (2012) presented a new theory of massive gravity and summarized that the fractional graviton masses of very low cosmic fluid density are different of zero. Shchigolev (2011, 2016) has reliably obtained the solutions for dynamical functional equations with fractional derivatives or based on the behavior with Einstein-Hilber. El-Nabulsi (2013) used fractional derivatives to generalize the Einstein's field equations and obtained non-local fractional Einstein's field equations. El-Nabulsi (2017) introduced a generalized derivative operator and derived a family of Emden–Fowler differential equations. El-Nabulsi (2016) used Ornstein-Uhlenbeck-like fractional differential equation and introduced a generalized fractional scale factor and a time-dependent Hubble parameter and described the accelerated expansion of a non-singular universe with and without the presence of scalar fields. Nouh (2019) solved the system of the fractional differential equations modeled the stellar helium burning phase and explained the possibility of the application of the fractional models during stellar model calculations.

The fractional Lane-Emden equations have been analysed by El-Nabulsi (2011) for the white dwarf stars, Bayian and Krisch (2015) for the incompressible gas sphere, Abdel-Salam and Nouh (2016) for the fractional isothermal gas sphere and Nouh and Abdel-Salam (2018) for the polytropic gas sphere, Abdel-Salam and Nouh (2020) for the conformable polytropic gas spheres. The fractional isothermal gas sphere has been analyzed by Mizra (2009) who solved the fractional isothermal gas spheres starting by two terms series expansion and iterates till obtaining reasonable values for the Emden function near the center of the sphere. By implementing Riemann-Liouville concept of the fractional derivatives, the fractional version of the equation is studied by Abdel-Salam and Nouh (2016). Abdel-Salam et al. (2020) developed a conformable Adomian decomposition method to obtain a divergent series solution for the isothermal gas spheres.

In the present article, we solve the conformable fractional isothermal gas sphere via the power series method and obtain a recurrence relation for the power series coefficient and comparing the current numerical results to previous studies. To evaluate the fractional parameter impact on the configuration of the stars, the physical parameters of the isothermal gas sphere are derived and determined for the neutron stars. The characteristic of the conformable derivatives motivates these calculations; it treats the continuous and discontinuous media and so the results may be theoretically important for the isothermal gas spheres.

The article is organized as follows. Section 2 describes the concepts of consistent fractional derivatives. Section 3 defines the series solution for the fractional isothermal gas sphere. The findings reported are in section 4 and the conclusion drawn is in section 5.

## 2. Conformable Fractional Derivatives

There are various definitions of fractional derivatives. Examples include Riemann–Liouville, Caputo, modified Riemann–Liouville, Kolwankar–Gangal, Cresson's, and Chen's fractal derivatives, Mainardi (2010) and Herrmann (2014).

Khalil et al. (2014) introduced the conformable fractional derivative (CFD) by using the limits in the form

$$D^{\alpha} f(t) = \lim_{\varepsilon \to 0} \frac{f(t + \varepsilon t^{1-\alpha}) - f(t)}{\varepsilon} \quad \forall t > 0, \ \alpha \in (0,1] \ , \tag{1}$$

$$f^{(\alpha)}(0) = \lim_{t \to 0^+} f^{(\alpha)}(t) \ . \tag{2}$$

Here $f^{(\alpha)}(0)$ is not defined. When $\alpha = 1$ this fractional derivative reduces to the ordinary derivative. The CFD has the following properties:

$$D^{\alpha} t^p = p t^{p-\alpha}, \ p \in \mathbb{R}, \ D^{\alpha} c = 0, \ \forall f(t) = c \ , \tag{3}$$

$$D^{\alpha}(af + bg) = a D^{\alpha} f + b D^{\alpha} g, \ \forall a,b \in \mathbb{R} \ , \tag{4}$$

$$D^\alpha(f\ g) = f\ D^\alpha g + f\ D^\alpha g\ , \qquad (5)$$

$$D^\alpha f(g) = \frac{df}{dg} D^\alpha g, \qquad D^\alpha f(t) = t^{1-\alpha} \frac{df}{dg}, \qquad (6)$$

where $f$, $g$ are two $\alpha-$differentiable functions and $C$ is an arbitrary constant. Equations (5) to (6) are proved by Khalil et al (2014). The CFD of some functions

$$D^\alpha e^{ct} = c t^{1-\alpha} e^{ct}, \quad D^\alpha \sin(ct) = c t^{1-\alpha} \cos(ct), \quad D^\alpha \cos(ct) = -c t^{1-\alpha} \sin(ct),$$
$$D^\alpha e^{ct^\alpha} = c \alpha e^{ct^\alpha}, \quad D^\alpha \sin(ct^\alpha) = c \alpha \cos(ct^\alpha), \quad D^\alpha \cos(ct^\alpha) = -c \alpha \sin(ct^\alpha). \qquad (7)$$

### 3. Fractional Equation of the Isothermal Gas Sphere

The polytropic equation of state has the form

$$p = K\rho, \qquad (8)$$

where $K$ is called the pressure constant. The equilibrium structure of a self-gravitating object is derived from the equations of hydrostatic equilibrium. The simplest case is that of a spherical, non-rotating, static configuration, where for a given equation of state all macroscopic properties are parameterized by a single parameter, for example, the central density.

The fractional form of equations of mass conservation and hydrostatic equilibrium is given by

$$\frac{d^\alpha M(r)}{dr^\alpha} = 4\pi r^{2\alpha} \rho, \qquad (9)$$

and

$$\frac{d^\alpha P(r)}{dr^\alpha} = -\frac{G M(r)}{r^{2\alpha}} \rho. \qquad (10)$$

Rearrange Equation (9) we get

$$\frac{r^{2\alpha}}{\rho} \frac{d^\alpha P(r)}{dr^\alpha} = -G M(r), \qquad (11)$$

By performing the first fractional derivative of Equation (14) we get

$$\frac{d^\alpha}{dr^\alpha}\left(\frac{r^{2\alpha}}{\rho} \frac{d^\alpha P(r)}{dr^\alpha}\right) = -G \frac{d^\alpha M(r)}{dr^\alpha}. \qquad (12)$$

Combining Equations (8) and (12) we get

$$\frac{d^\alpha}{dr^\alpha}\left(\frac{r^{2\alpha}}{\rho}\frac{d^\alpha P(r)}{dr^\alpha}\right) = -4\pi G r^{2\alpha}\rho, \tag{13}$$

or

$$\frac{1}{r^{2\alpha}}\frac{d^\alpha}{dr^\alpha}\left(\frac{r^{2\alpha}}{\rho}\frac{d^\alpha P(r)}{dr^\alpha}\right) = -4\pi G \rho. \tag{14}$$

Now, by defining the dimensionless function $u$ (Emden function) as

$$\rho = \rho_c e^{-u}, \tag{15}$$

where $\rho$ and $\rho_c$ are the density and central density respectively. The dimensionless variable $x$ could be written as

$$x^\alpha = \frac{r^\alpha}{a}. \tag{16}$$

Inserting Equations (8) and (15) in Equation (14) we get

$$\frac{1}{(ax^\alpha)^2}\frac{d^\alpha}{d(ax^\alpha)}\left(\frac{(ax^\alpha)^2}{\rho_c e^{-u}}\frac{d^\alpha(K\rho)}{d(ax^\alpha)}\right) = -4\pi G \rho_c e^{-u}, \tag{17}$$

The fractional derivative of the Emden function $e^{-u}$ could be written as

$$\frac{d^\alpha}{dx^\alpha}e^{-u} = -e^{-u}\frac{d^\alpha u}{dx^\alpha}. \tag{18}$$

Inserting Equation (18) in Equation (17) we get

$$\frac{K}{a^2 x^{2\alpha}}\frac{d^\alpha}{dx^\alpha}\left(\frac{-x^{2\alpha}\rho_c e^{-u}}{\rho_c e^{-u}}\frac{d^\alpha u}{dx^\alpha}\right) = -4\pi G \rho_c e^{-u}, \tag{19}$$

rearrange

$$\frac{K}{4\pi G \rho_c a^2}\frac{1}{x^{2\alpha}}\frac{d^\alpha}{dx^\alpha}\left(x^{2\alpha}\frac{d^\alpha u}{dx^\alpha}\right) = e^{-u}. \tag{20}$$

Now by taking

$$a^2 = \frac{K\rho_c^{-1}}{4\pi G}. \tag{21}$$

Therefore the fractional isothermal Lane-Emden equation in its fractional form is given by

$$\frac{1}{x^{2\alpha}}\frac{d^\alpha}{dx^\alpha}\left(x^{2\alpha}\frac{d^\alpha u}{dx^\alpha}\right) = e^{-u}. \tag{22}$$

## 4. Physical Characteristic of the Fractional Isothermal Ga Sphere

The mass contained in a radius $r$ is given by

$$M(r^\alpha) = \int_0^r 4\pi r^{2\alpha} \rho \, dr^\alpha .  \tag{23}$$

Inserting Equations (15) and (16) for $\rho$ and $r^\alpha$ we found

$$M(x^\alpha) = 4\pi a^3 \rho_c \int_0^x x^{2\alpha} e^{-u} \, dx^\alpha ,  \tag{24}$$

by substituting Equation (22) for the Emden function $e^{-u}$ we get

$$M(x^\alpha) = 4\pi a^3 \rho_c \int_0^x x^{2\alpha} \left[ \frac{1}{x^{2\alpha}} \frac{d^\alpha}{dx^\alpha} \left( x^{2\alpha} \frac{d^\alpha u}{dx^\alpha} \right) \right] dx^\alpha$$

$$= 4\pi a^3 \rho_c \int_0^x \left[ \frac{d^\alpha}{dx^\alpha} \left( x^{2\alpha} \frac{d^\alpha u}{dx^\alpha} \right) \right] dx^\alpha$$

$$= 4\pi a^3 \rho_c \left[ \left( x^{2\alpha} \frac{d^\alpha u}{dx^\alpha} \right) \right],  \tag{25}$$

with $a$ could be followed from Equation (21), then the mass is given by

$$M(x^\alpha) = 4\pi \left[ \frac{K}{4\pi G} \right]^{\frac{3}{2}} \rho_c^{-\frac{3}{2}} \left[ \left( x^{2\alpha} \frac{d^\alpha u}{dx^\alpha} \right) \right].  \tag{26}$$

The radius of the isothermal gas sphere is given by

$$R^\alpha = a x^\alpha .  \tag{27}$$

Inserting $a$ in Equation (27) we get

$$R^\alpha = \left[ \frac{K}{4\pi G} \right]^{\frac{1}{2}} \rho_c^{-\frac{1}{2}} x^\alpha .  \tag{28}$$

The pressure and temperature of the polytrope are given by

$$P = P_c \, e^{-u},  \tag{29}$$

and

the central density is given by

$$cc = \frac{\rho_c}{\bar{\rho}} = \frac{x_1}{3\left[\left(\dfrac{d^\alpha u}{d x^\alpha}\right)\right]_{x=x1}} \tag{30}$$

where $\bar{\rho}$ is the mean density and $x_1=80$ in the present calculations.

## 5. Series Solution of Equation (22)

Equation (22) could be written as

$$x^{2\alpha} D_x^\alpha D_x^\alpha u + 2\alpha^2 x^\alpha D_x^\alpha u + x^{2\alpha} e^{-u} = 0, \qquad u(0)=0, \quad D_x^\alpha u(0)=0 \tag{31}$$

where $u = u(x)$ is an unknown function and $\alpha \in (0,1]$. We assume the transform $X = x^\alpha$ and the solution can be expressed in a series in the form

$$u(X) = \sum_{m=0}^{\infty} A_m X^m = A_0 + A_1 x^\alpha + A_2 x^{2\alpha} + A_3 x^{3\alpha} + A_4 x^{4\alpha} + A_5 x^{5\alpha} + \dots, \tag{32}$$

At x=0 we have

$u(0) = A_0 = 1$,

$D_x^\alpha u(0) = 0$,

$A_1 = 0$.

Now the term $e^{-u}$ could be expressed as

$$e^{-u(X)} = G(X), \tag{33}$$

where

$$G(X) = \sum_{m=0}^{\infty} Q_m X^m = Q_0 + Q_1 X + Q_2 X^2 + Q_3 X^3 + Q_4 X^4 + Q_5 X^5 + \dots , \tag{34}$$

At x=0 we get

$G(0) = Q_0 = 1$.

Performing the fractional derivative $G$, we obtain

$D_x^\alpha G = D_x^\alpha e^{-u}$,

$-e^{-u} D_x^\alpha u = D_x^\alpha G$,

$$G D_x^\alpha u = -D_x^\alpha G . \tag{35}$$

By having the k-times differentiation on both sides of equation (35), we can obtain

$$\underbrace{D_x^\alpha ... D_x^\alpha}_{k \text{ times}} [GD_x^\alpha u] = -\underbrace{D_x^\alpha ... D_x^\alpha}_{k \text{ times}} (D_x^\alpha G) \Rightarrow \sum_{j=0}^{k} \binom{k}{j} u^{(\alpha(j+1))} G^{(\alpha(k-j))} = -G^{(\alpha(k+1))}, \tag{36}$$

From the last equation, we get

$$\sum_{j=0}^{k} \binom{k}{j} (j+1)! \alpha^{j+1} (k-j)! \alpha^{k-j} Q_{k-j} A_{j+1} = -(k+1)! \alpha^{k+1} Q_{k+1},$$

After some simplifications, we obtained

$$Q_{k+1} = -\frac{1}{(k+1)! \alpha^{k+1}} \sum_{j=0}^{k} k!(j+1) \alpha^{k+1} Q_{k-j} A_{j+1},$$

let $l = k+1$ and $i = j+1$, then we have

$$Q_l = -\frac{(l-1)!}{l! \alpha^l} \sum_{i=1}^{l} i \alpha^l Q_{l-i} A_i = -\frac{1}{l} \sum_{i=1}^{l} i Q_{l-i} A_i. \tag{37}$$

Now we turn to determine the recurrence relation of the series expansion of the Emden function $u$. By taking the first and second derivatives of Equation (32), we get

$$D_x^\alpha u = \sum_{m=1}^{\infty} A_m m \alpha X^{m-1} x^{\alpha-\alpha} = \sum_{m=1}^{\infty} A_m m \alpha X^{m-1},$$

and

$$D_x^\alpha D_x^\alpha u = \sum_{m=2}^{\infty} A_m m(m-1) \alpha^2 X^{m-2}.$$

Substituting the two last equations into Equation (31) yield

$$x^{2\alpha} \sum_{m=2}^{\infty} A_m m(m-1) \alpha^2 X^{m-2} + 2\alpha x^\alpha \sum_{m=2}^{\infty} A_m m \alpha X^{m-1} - x^{2\alpha} \left[ 1 + \sum_{m=1}^{\infty} Q_m X^m \right] = 0,$$

Rearranging terms we get

$$\sum_{m=0}^{\infty} A_{m+2}(m+2)(m+1) \alpha^2 X^{m+2} + \sum_{m=0}^{\infty} 2\alpha^2 (m+2) A_{m+2} X^{m+2} - \left[ X^2 + \sum_{m=1}^{\infty} Q_m X^{m+2} \right] = 0.$$

The series coefficients $A_{m+2}$ are obtained by equating the coefficients of the $X^{m+2}$ as

$$A_{m+2} = \frac{Q_m}{\alpha^2 (m+2)[(m+1)+2]}. \tag{38}$$

From the coefficients of the $X^2$ term, we get

$2\alpha^2 A_2 + 4\alpha^2 A_2 - 1 = 0$,

$A_2 = \dfrac{1}{6\alpha^2}$.

From the $A_2$ and Equation (38), we get

$Q_2 = -\dfrac{1}{6\alpha^2}$.

The rest of the coefficients of the series expansion could be obtained using the two recurrence relations Equations (37-38). The Emden function of the isothermal gas sphere represented by the first six terms of the series expansion is given by

$$u(x) = \dfrac{x^2}{6\alpha^2} - \dfrac{x^4}{120\alpha^4} + \dfrac{x^6}{1890\alpha^6} - - -, \tag{39}$$

where $A_0 = A_1 = A_3 = 0$.

## 6. Results

The structure of the conformable isothermal gas sphere is determined by the analytical solution of Equation (32) with the two recurrence relations, Equations (37-38). A comparison between the original series solution (the curve labeled *a*) computed at $\alpha = 1$ and the numerical solution (the curve labeled *b*) is illustrated in Figure 1. It is worthy noted that without applying acceleration techniques the Emden function computed using the power series is constrained. The power series solution is slowly convergent and completely divergent beyond $x \simeq 3.2$. Therefore, the power series solution for the isothermal gas spheres must be accelerated to reach the surface of the spheres accurately, and consequently, effects help to achieve a better physical approximation of the stars. To do that, a combination of the two techniques for the transformation of Euler-Abel and the approximation of the Pad'e was employed (Nouh, 2004; Nouh & Saad 2013). Table 1 lists the comparison between the Emden function ($u_a$) computed using the accelerated series expansion and the Emden function ($u_n$) representing the numerical solution by (Horedt, 1987). The maximum relative error is ~2%.

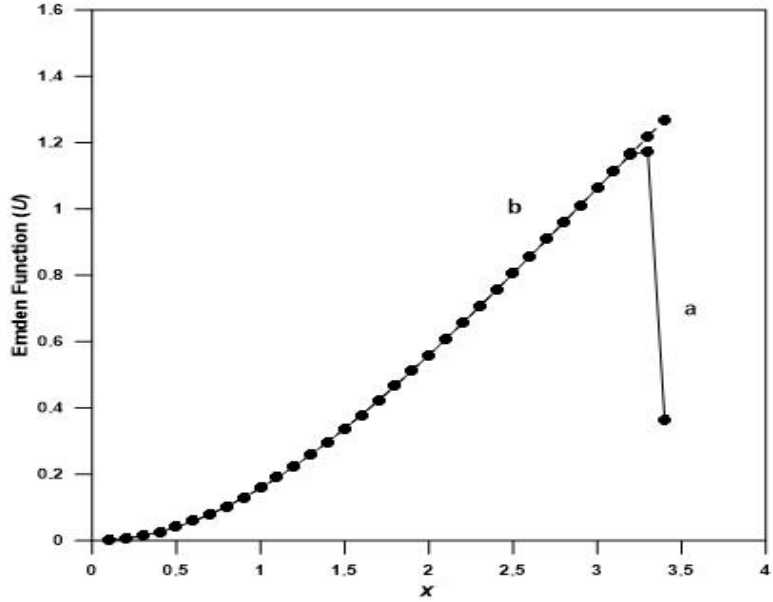

Figure 1: Comparison between the series solution before applying the acceleration technique and the numerical solution for the integer case $\alpha = 1$.

Table 1: Comparison between the accelerated integer Emden function ($u_a$) and the corresponding numerical one ($u_n$).

| X | $u_a$ | $u_n$ |
|---|---|---|
| 1 | 0.1588 | 0.1588 |
| 5 | 2.0440 | 2.0440 |
| 10 | 3.7365 | 3.7365 |
| 15 | 4.7322 | 4.7341 |
| 20 | 5.3902 | 5.4038 |
| 30 | 6.2831 | 6.2840 |
| 40 | 6.8602 | 6.8677 |
| 45 | 7.0834 | 7.0988 |
| 50 | 7.2750 | 7.3022 |
| 60 | 7.5846 | 7.6484 |
| 70 | 7.8200 | 7.9368 |
| 80 | 8.0016 | 8.1844 |

Now we turn to study the effects of the fractional parameters on the Emden function. Figure

2 illustrates the accelerated Emden function $u_a$ (solid lines) versus the dimensionless parameter $x$, the calculations are performed for the fractional parameters $\alpha = 0.8 - 1$ with step $\Delta\alpha = 0.05$. As we can notice from the figure and Equation (39) that, $u_a(x)$ is increased by decreasing the fractional parameters $\alpha$. This behavior is not the same as the fractional polytropic gas sphere obtained by Nouh and Abdel-Salam (2018) and Abdel-Salam and Nouh (2020).

The fractional Emden functions ($u_R$) computed by Abdel-Salam and Nouh (2016) are plotted (dashed lines) along with the conformable one ($u_c$) in Figure 2. The calculations were performed in the sense of the modified Rieman-Liouville fractional derivatives (mRFD). We can notice the difference between the two fractional functions; while the Emden function $u_c$ increases by increasing the dimensionless distance, the Emden function $u_R$ is nearly constant for the dimensionless distance $x \simeq 35$. This may come from the difference in the nature between the conformable and the mRFD. The conformable derivative is applicable for differentiable and non-differentiable functions or applied in continuous and non-continuous media but mRL derivative is applicable for non-differentiable functions or non-continuous media only.

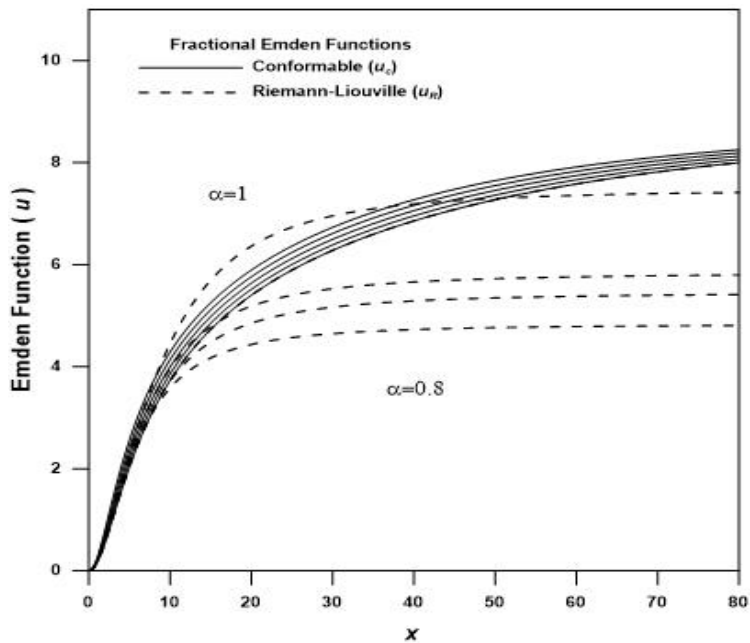

Figure 2: The fractional Emden functions computed by the conformable derivatives (solid lines) and that computed by Rieman-Liouville fractional derivatives (dashed lines) for different values of the fractional parameter $\alpha$.

Based on om Equations (26-30), we computed and examined the structure of the neutron stars, such as radius-mass relationship, density ratio, and pressure ratio. In this context, we applied the accelerated analytical solution in the present paper to the neutron star with the physical parameters: mass $M = 1.4 M_\odot$, central density $\rho_c = 5.75 \times 10^{14}$ g cm$^{-3}$, pressure $P = 2 \times 10^{33}$ par, and radius $R = 1.4 \times 10^6$ cm. The results for the radii, masses, and central density of the conformable models are listed in Table 2. We observed that the conformable fractional models have less volume and less mass than both the integer models and the fractional models computed based on the scheme presented by Abdel-Salam and Nouh (2016). The dramatic change in both the mass and radius of the models indicate the huge compression of the material inside the star. This is very clear from the great changes in the values of the central condensation computed by Equation (30) and listed in the fourth column of Table 2. Figures (3-4) plot the distributions of the fractional masses and densities versus the stellar radius at various fractional parameters.

Table 2: Mass-Radius relation of the conformable isothermal gas spheres.

| $\alpha$ | $R/R_0$ | $M/M_0$ | cc |
|---|---|---|---|
| 1 | 0.998 | 0.998 | 1698.4 |
| 0.95 | 0.802 | 0.644 | 1142.6 |
| 0.9 | 0.644 | 0.415 | 771.2 |
| 0.85 | 0.517 | 0.268 | 566.6 |
| 0.8 | 0.415 | 0.173 | 350.7 |
| 0.75 | 0.334 | 0.111 | 236.0 |

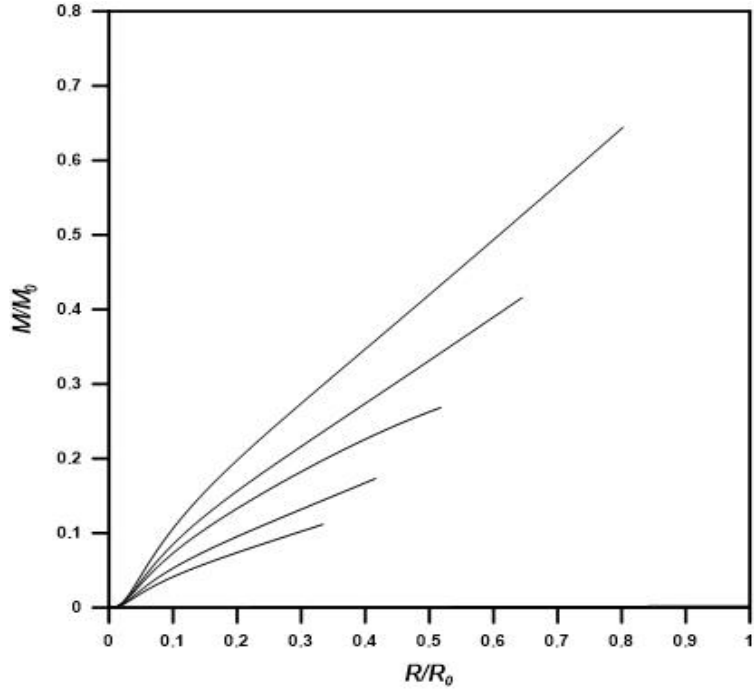

Figure 3: Conformable fractional mass-radius relations for a typical neutron star.

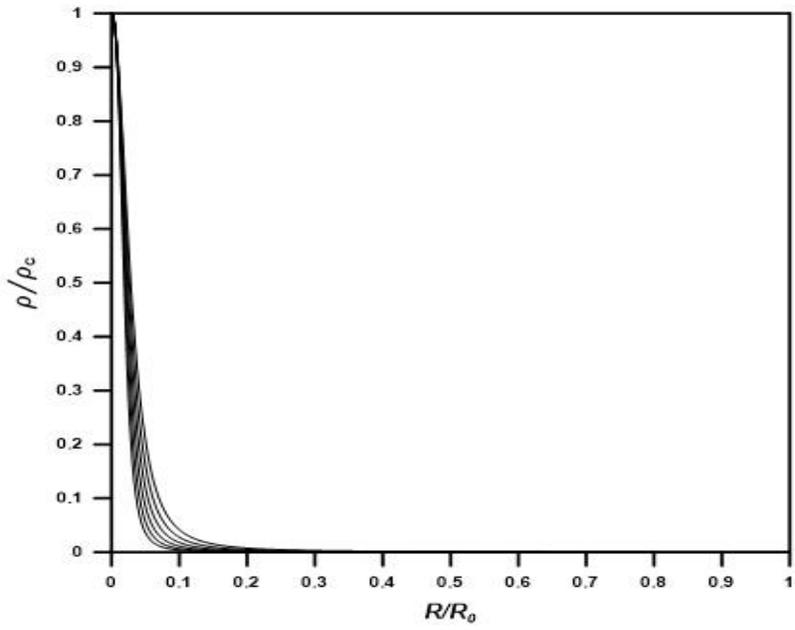

Figure 4: The distribution of the density ratio for the conformable fractional models of neutron stars.

## 7. Conclusion

We have developed general analytical formulations in the context of fractional Newtonian isothermal gas spheres. Besides the Isothermal gas sphere equation, we used the conformable fractional calculus to obtain analytical expressions for the mass, radius, and density distributions. Based on the Series expansion method, we obtained recurrence relation which enables us to calculate the series coefficients. For the fractional parameter $\alpha = 1$, the numerical comparison between the Emden functions determined by the series solution and those calculated using the Runge-Kutta method is good for the dimensionless parameters up to $x \simeq 3.2$ and after that, the series diverges.

To improve the convergence behavior of the series we used the scheme developed by Nouh (2004) and Nouh and Saad (2013) to accelerate the series. Then the accelerated series is converged to the probable values calculated by the numerical integration. Using the accelerated series expansion, we calculated the Emden functions for different fractional parameters. As appeared from Equation (39) and Figure 2 the Emden function increased by decreasing the fractional parameter.

We compared the Emden function computed in the present article with that computed by Abdel-Salam and Nouh (2016). The difference in the behavior of the two functions is attributed to the difference between the CFD and mRFD. While the conformable derivative could be applied for differentiable and non-differentiable functions. So, CFD could be applied for continuous and discontinuous media, it is not the case for the mRFD.

Calculation and study are carried out of the fractional physical parameters such as the mass-radius relation, density, and pressure ratios for different fractional models for the neutron star structure were performed. We found that the existing models of the conformable fractional stars have less volume and mass than the integer models.